# Estimation of incidence from aggregated current status data

Ralph Brinks

Chair for Medical Biometry and Epidemiology, Witten/Herdecke University, Faculty of Health/School of Medicine, D-58448 Witten, Germany

Institute for Biometry and Epidemiology, German Diabetes Center, D-40225 Düsseldorf, Germany

Corresponding author:
E-Mail: ralph.brinks@uni-wh.de

## Abstract

We use historical data about breathlessness in British coal miners and recent data about diabetes in Germany to illustrate a method for deriving the age-specific incidence from aggregated current status data, i.e. age-specific prevalence data. The method is centered on a differential equation, and special focus is put on maximum likelihood (ML) estimation of confidence intervals.

Key words: prevalence, chronic diseases, illness-death model, maximum likelihood estimation

# Introduction

The question about relations between incidence and prevalence of diseases dates back at least to 1934, when Muench examined if the age-specific incidence of yellow fever in Southern America can be estimated from a cross-sectional sample about the age-specific prevalence [Mue34]. Muench used the terms *summation data* and *catalytic curve*, the latter still being found in modern textbooks about infectious disease epidemiology [Vyn10]. In the statistical literature, the term *aggregated current status data* is more frequently used than *summation data*. For current status data, it is assumed to have a sample where for each study participant, (mostly) a binary disease status $Y$ and a time variable $T$ are given. The time variable $T$ describes the point in time the disease status $Y$ refers to. For example, consider the situation when a diagnostic test for detecting antibodies of SARS-CoV-2 (where $Y$ describes whether antibodies are present or not) is applied to population samples during different phases ($T$) of the pandemic. A review about current status data is found in [Jew03, Jew04]. For a textbook introduction about current status data, we refer to [Sun06]. In epidemiology, current status data are often aggregated in different age groups. For example, the data in Muench's article about yellow fever is given as positive cases ($c$) from a number of tests ($n$) applied in five age groups: 5-9, 10-14, 15-19, 20-39 and 40+ (Table 1 in [Mue34]).

In this article, we briefly review the illness-death model (IDM) for chronic conditions and its relation to current status data. Aggregated current status data can be expressed in terms of the prevalence of the chronic condition, and we focus on mathematical relations between the age-specific prevalence and the transition rates in the IDM. As such, the article follows the tradition of Keiding's review article [Kei91]. We, however, focus on mathematical relations that are based on differential equations similar to the Kolmogorov Forward Equations [Bri18]. The differential equations are then combined with maximum-likelihood estimation to obtain the incidence from the aggregated current status data. Using the mathematical relations for estimating incidence from prevalence data has been described in [Bri13] and has been compared to other methods in [Lan16]. So far, for inferential statistics we have used re-sampling techniques, see e.g. [Bri15]. In this work, we use a maximum likelihood approach and put special emphasis on estimating confidence intervals.

As applications, we use a historical dataset about breathlessness in British coal miners [Ela14] and recent data about type 2 diabetes in Germany [Tam16] to estimate the age-specific incidence from aggregated current status data.

# Illness death model

Recently, it has been shown that the age-specific prevalence $p$ of a chronic condition at some time $t$ is related to the age-specific incidence density $i$ (synonym: incidence rate), and mortality rates $m_0$ and $m_1$ via a partial differential equation (PDE) [Bri14, Bri15a]. Figure 1 shows the underlying multi-state model with the possible transitions and associated rates. The prevalence $p$ and the rates $i$, $m_0$, and $m_1$ in the IDM generally depend on the calendar time $t$ and on the age $a$. For instance, $p(t, a)$ and $i(t, a)$ denote the fraction of people alive with the condition and the incidence rate of the people aged $a$ at time $t$, respectively. In epidemiological contexts, the calendar time $t$ is sometimes called period [Cla87].

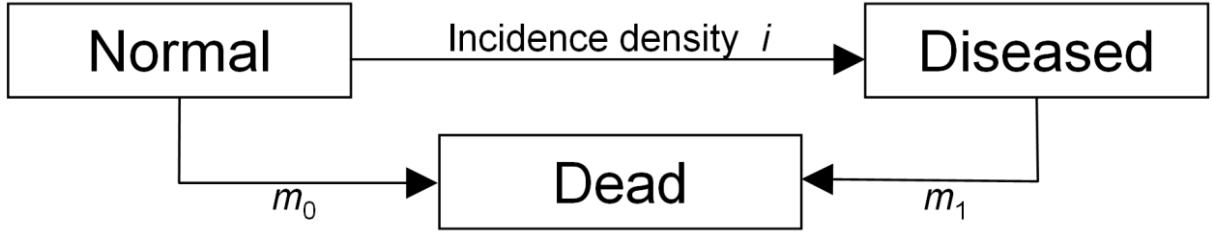

**Figure 1: Illness-death model for a chronic condition ('Diseased') and associated transition densities: incidence density $i$, mortality without ($m_0$) and with the disease ($m_1$).**

The PDE linking the transition rates $i$, $m_0$, and $m_1$ from the IDM in Figure 1 with the age-specific prevalence $p$ reads as

$$(\partial p/\partial t + \partial p/\partial a) = (1 - p)\{i - p(m_1 - m_0)\}, \tag{1}$$

where $\partial/\partial x$ means the partial derivative with respect to $x$.

## Maximum likelihood estimation

Assume that we have aggregated current status data for age groups indexed $k$, $k = 1, ..., K$ and that age group $k$ has reported $c_k$ people with the chronic condition from an overall number $n_k$ of people in that age group. Then, the binomial likelihood function $L$ for the aggregated current status data is given by

$$L = \prod_{k=1}^{K} \binom{n_k}{c_k} p_k^{c_k} (1 - p_k)^{n_k - c_k} \tag{2}$$

where $k$ is the index for the age group and $p_k$ is the (unknown) prevalence, i.e., the fraction of subjects in age group $k$ with the condition under consideration.

In case the prevalences $p_k$, $k = 1, ..., K$, in Equation (2) have an analytical representation as a function of the rates $i$, $m_0$, and $m_1$, i.e., $p = p(i, m_0, m_1)$, estimation of parameters can be straightforward, which we will demonstrate in data about breathlessness in British coal miners. The observed aggregated current status data about breathlessness are shown in Table 1. To incorporate differences in mortality between diseased and non-diseased people, we will also use mortality data (Table 2).

Usually, mortality rates $m_0$ and $m_1$ are not available in epidemiological contexts. It is more common to have the general mortality $m$ and the mortality rate ratio $R = m_1/m_0$ available. In this situation, the prevalence $p$ does not have an analytical representation $p = p(i, m, R)$ which will be explained later. Then, the incidence can be estimated by a plug-in estimator, which will be demonstrated in data about type 2 diabetes shown in Table 3.

## Data sets

Table 1 shows the observed aggregated current status data about breathlessness in British coal miners [Ela14].

| Age group $k$ (in years) | Number observed ($n_k$) | Number with condition ($c_k$) |
|---|---|---|
| 20-24 | 1952 | 16 |
| 25-29 | 1791 | 32 |
| 30-34 | 2113 | 73 |
| 35-39 | 2783 | 169 |
| 40-44 | 2274 | 223 |
| 45-49 | 2393 | 357 |
| 50-54 | 2090 | 521 |
| 55-59 | 1750 | 558 |
| 60-64 | 1136 | 478 |

**Table 1: Data about breathlessness in British coal miners (Table 14.2a in [Ela14]).**

Later, we are also interested in incorporating mortality. Hence, we use the life tables also presented in [Ela14].

| Age group $k$ (in years) | Life tables | |
| | General population | Population with breathlessness |
|---|---|---|
| 20-24 | 481185 | 343937 |
| 25-29 | 478683 | 333343 |
| 30-34 | 476150 | 320446 |
| 35-39 | 472641 | 304305 |
| 40-44 | 467066 | 284325 |
| 45-49 | 457729 | 260806 |
| 50-54 | 441895 | 233060 |
| 55-59 | 415262 | 200561 |
| 60-64 | 372908 | 163241 |

**Table 2: Life tables for England and Wales (Table 14.2b in [Ela14]).**

To demonstrate an alternative way to estimate incidence based on a plug-in estimator, we use the data shown in Table 3, which shows aggregated current status data about ascertained diagnosis of type 2 diabetes of women from the German statutory health insurance in the years 2009 and 2010. Details about the data are described in [Tam16]. Due to strict restrictions in the use of the original data, random noise (2%) has been added to the original data; then data have been downsampled by factor of 100 and rounded to the nearest integer.

| Age group $k$ (in years) | Women year 2009 | | Women year 2010 | |
| | Observed ($n_k$) | With diabetes ($c_k$) | Observed ($n_k$) | With diabetes ($c_k$) |
|---|---|---|---|---|
| 20-24 | 19029 | 33 | 18939 | 36 |
| 25-29 | 19549 | 65 | 18917 | 69 |
| 30-34 | 19391 | 109 | 19388 | 122 |
| 35-39 | 20885 | 200 | 19722 | 208 |
| 40-44 | 28844 | 402 | 26543 | 403 |
| 45-49 | 28856 | 706 | 29509 | 742 |
| 50-54 | 25641 | 1145 | 25870 | 1178 |
| 55-59 | 23223 | 1826 | 23238 | 1850 |
| 60-64 | 18845 | 2134 | 20112 | 2423 |
| 65-69 | 21964 | 3160 | 19714 | 2887 |
| 70-74 | 22965 | 4281 | 23452 | 4446 |
| 75-79 | 15944 | 3628 | 16509 | 3909 |
| 80-84 | 13310 | 3114 | 13083 | 3295 |
| 85-89 | 8796 | 2159 | 8637 | 2220 |
| 90-94 | 2380 | 569 | 2760 | 681 |
| 95-99 | 892 | 188 | 833 | 177 |

**Table 3: Diabetes in German women from the statutory health insurance 2009 and 2010 [Tam16].**

# Estimation of the age-specific incidence

We distinguish the situation that people with and without the disease have a different mortality or not. The latter is called non-differential mortality [Kei91].

## *Non-differential mortality*

First, we consider the case without migration and without differential mortality (i.e., $m_0 = m_1$), when the PDE (1) becomes

$$(\partial p/\partial t + \partial p/\partial a) = (1 - p)\, i. \qquad (3)$$

If we assume furthermore, that the prevalence $p$ in Equation (1) does not depend on time $t$, the PDE (1) becomes an ordinary differential equation (ODE)

$$\partial_a p := dp/da = (1-p)\, i, \qquad (4)$$

which has the general solution

$$p(a) = 1 - (1 - p_0) \times \exp\left(-\int_{a_0}^{a} i(\tau)d\tau\right) \qquad (5)$$

where $p_0 = p(a_0)$ is the initial condition.

Equation (4) is the basis for a straightforward estimator of the age-specific incidence density $i$. If we can estimate the derivative $\partial_a p$ from the age-specific prevalence $p$, we obtain $i = \partial_a p /(1 - p)$.

Example 1: We fit a linear regression model with logit-transformed prevalence as outcome $\text{logit}(p(a)) = \log(p(a)/\{1 - p(a)\}) = \beta_0 + \beta_1 \times a$ at the midpoints of the age groups $a = 22.5, 27.5, ..., 62.5$ to the data from Table 1. We obtain $\beta_0 = -7.02$ and $\beta_1 = 0.110$. Inserting $p(a) = \text{expit}(\beta_0 + \beta_1\, a)$ into $i = \partial_a p/(1 - p)$ yields $i(a) = \beta_1\, p(a)$. The expit-function is the inverse of the logit-function, i.e., $(\text{logit})^{-1} = \text{expit} = \exp/(1+\exp)$, and has the derivative $\text{expit}´ = \text{expit} \cdot (1 - \text{expit})$. The associated age-specific incidence $i(a) = \beta_1\, \text{expit}(\beta_0 + \beta_1\, a)$ with $\beta_0 = -7.02$ and $\beta_1 = 0.110$ is plotted as black line in Figure 2. For comparison, the blue line shows the estimates [Ela14] has given. Note the negative estimate at age 57.5 from [Ela14].

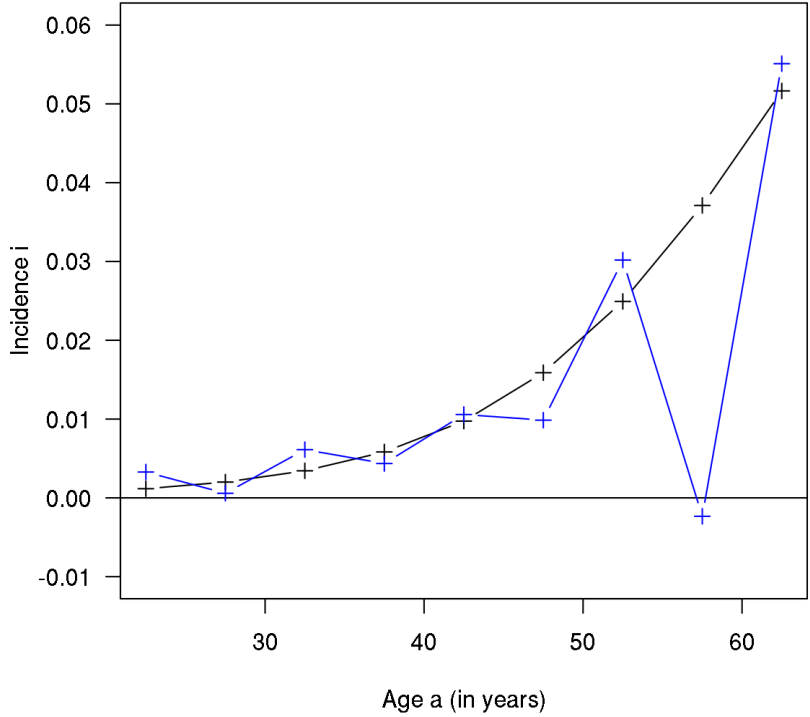

**Figure 2: Estimated age-specific incidence for the data in Table 1 as calculated in Example 1.**

Example 1 has demonstrated how the theory of differential equations can be used to estimate the age-specific incidence from aggregated current status data of a chronic condition. Since we are interested in additionally estimating confidence intervals for the age-specific incidence, we chose the maximum likelihood approach.

With a view to Figure 2, we have hints that the age-specific incidence density $i(a)$ grows exponentially with age. Hence, we make the approach $i(a) = \exp(\beta_0 + \beta_1 \cdot a)$ with coefficients $\beta_0, \beta_1$. If we substitute this incidence into Equation (5) with the initial condition $a_0 = 20$ and $p(20) = p_0 = 0$, we obtain $p(a) = 1 - \exp(h(20) - (a))$ with $h(z) = \exp(\beta_0 + \beta_1 \cdot z)/\beta_1$. Then, this $p$ is evaluated at the age group midpoints $a_k = 22.5, \ldots, 62.5$ and finally substituted into Equation (2). We end up at a likelihood function $L = L(\beta_0, \beta_1)$, which by optimization can be used to calculate a maximum likelihood estimator for both coefficients, $\beta_0$ and $\beta_1$. If we additionally calculate the 95% confidence intervals based on the standard error derived from the inverse of the Fisher information matrix for large sample approximation of the variance-covariance matrix [Woo15], we obtain the results shown in Table 3.

|            | Point estimate | 95% confidence intervals |
|------------|----------------|--------------------------|
| $\beta_0$  | −7.8237        | −8.0590 to −7.5883       |
| $\beta_1$  | 0.07559        | 0.07006 to 0.08111       |

**Table 4: Maximum likelihood estimators for the coefficients $\beta_0$ and $\beta_1$ used for parameterization of the age-specific incidence without differential mortality.**

## Differential mortality

In case of differential mortality, we have to distinguish the situations which type of mortality is given. In the next example, we have the mortality rate ($m_1$) of diseased subjects available and the general mortality ($m$) in the overall population. After that, we consider the case where the general mortality $m$ and the mortality rate ratio $R = m_1/m_0$ are available.

### Mortality of diseased and general mortality

Starting from Equation (1) and using the fact that the general mortality rate $m$ is a convex combination of $m_0$ and $m_1$, $m = (1 - p)\, m_0 + p\, m_1$, we obtain

$$(\partial/\partial t + \partial/\partial a)\, p = i - p(m_1 - m + i). \tag{6}$$

Again, we assume that Equation (6) does not depend on time $t$ and obtain the ODE

$$\partial_a p = i - p\,(m_1 - m + i) \tag{7}$$

which has the general solution

$$p(a) = \exp(-G(a)) \times \left\{ p_0 + \int_{a_0}^{a} i(\tau) \times \exp(G(\tau))\, d\tau \right\} \tag{8}$$

with $G(a) = \int_{a_0}^{a} \{m_1 - m + i\}(\tau)\, d\tau$ and $p_0 = p(a_0)$ being the initial condition.

The general mortality rate $m$ and the mortality rate $m_1$ of the miners with breathlessness are estimated from the life tables shown in Table 2. This is done by converting the five-year probabilities of dying in the second and third columns of Table 2 an annual mortality rates $m$ and $m_1$ using the theory of single decrement processes (see e.g., Chapter 3 in [Pre01]).

Similar to the previous section, we make the approach $i(a) = \exp(\beta_0 + \beta_1 \cdot a)$ with coefficients $\beta_0$, $\beta_1$. If we substitute this incidence into Equation (8) with the initial condition $a_0 = 20$ and $p(20) = p_0 = 0$. The maximum likelihood estimator for both coefficients, $\beta_0$ and $\beta_1$, for the likelihood in Eq. (2) yields the results shown in Table 4.

|  | Point estimate | 95% confidence intervals |
|---|---|---|
| $\beta_0$ | –8.4706 | –8.7296 to –8.2116 |
| $\beta_1$ | 0.10107 | 0.094791 to 0.10734 |

**Table 5: Maximum likelihood estimators for the coefficients $\beta_0$ and $\beta_1$ used for parameterization of the age-specific incidence with differential mortality.**

### General mortality and mortality rate ratio

Given the data in Table 3, we make the approach $p(t, a) = \text{expit}(\beta_0 + \beta_1 \cdot t + \beta_2 \cdot a + \beta_3 \cdot a^2)$ and calculate the ML estimator for the coefficients $\beta_0, ..., \beta_3$. We assume that the general mortality $m$ of German women is known. Furthermore, the mortality rate ratio $R = m_1/m_0$ is assumed to be known. Given $m = (1 - p)\, m_0 + p\, m_1$ and $R = m_1/m_0$, the PDE (1) reads as

$$(\partial/\partial t + \partial/\partial a)p = (1-p)\left\{i - m\frac{p(R-1)}{1+p(R-1)}\right\}, \qquad (9)$$

which can be solved for the incidence

$$i = \frac{(\partial/\partial t + \partial/\partial a)p}{1-p} + m\frac{p(R-1)}{1+p(R-1)} \qquad (10)$$

For *m* and *R*, we take the values from the German Federal Statistical Office and the Danish Diabetes Register [Car19], respectively.

Once the ML estimate for *p* is found, Eq (10) yields the ML estimate for the incidence *i* via the plugin-principle [Was04]. Confidence intervals for *i* are calculated by using the delta method to transform the standard errors in *p* to those in *i* [Sha03]. The estimates of the age-specific incidence rate including the 95% confidence intervals are shown in Figure 3. At older ages, the length of the confidence intervals increases.

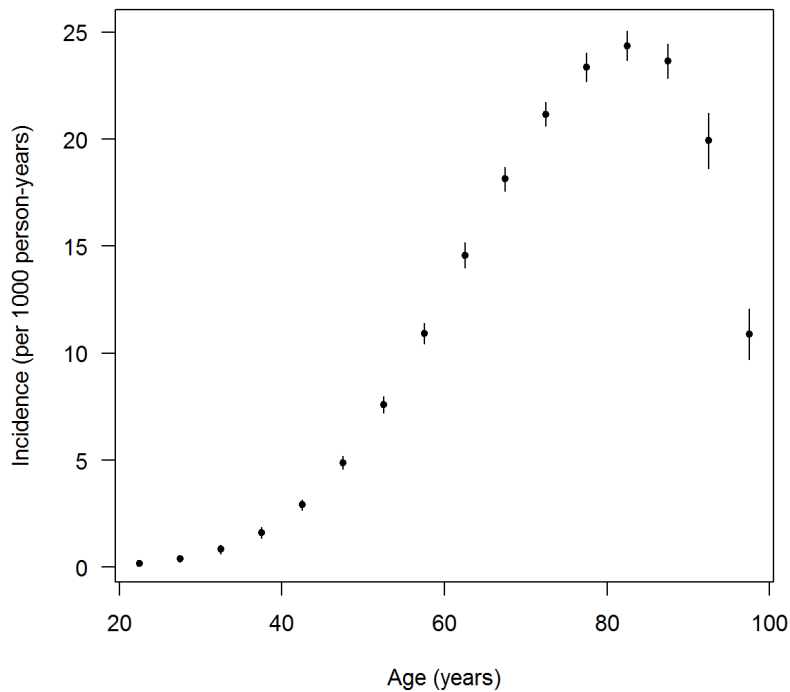

**Figure 3: Age-specific incidence rate of diabetes in women as estimated with the plug-in estimate Eq. (9). The vertical bars indicate the 95% confidence bounds.**

## Discussion

In this work, we have described a novel method about statistical inference around a differential equation, which relates the age-specific prevalence of a chronic condition with the underlying age-specific incidence. The method allows estimating the age-specific incidence rate from aggregated current status data (age-specific prevalence data). The roots of the differential equation go back at least to [Bru99]. Apart from estimating incidence rates, the differential equation has been used in other applications, e.g. mortality from prevalence and incidence [Toe18], in making projections about people with chronic conditions [Voe22] or estimating the effect of health policies [Toe21]. A historically interesting, epidemiological application is estimation of incidence from prevalence. In case of differential mortality, this can be accomplished if the mortality rate in the general population and the mortality rate ratio of people with disease over people without disease are given. General mortality may come from nationwide statistical offices, mortality rate ratios can be transferred from other settings, e.g., from other countries. This can be done, because rate ratios provide a stable measure of association in a wide variety of human populations [Bre80]. So far, there has not been a theory about statistical inference about the differential equation and we had to employ re-sampling techniques to obtain confidence bounds. In re-sampling, a number of random samples from the reported distributions of the input parameters have been drawn to estimate how the uncertainty in the input parameters propagates through the differential equations into the outcomes. An example for this type of re-sampling is described, e.g., in [Bri15].

The question arises if and how the methods described in this paper can be generalized. In the examples with differential mortality, one might consider the situation where not only the prevalence but also the information about mortality are only known with statistical uncertainty. In this case, the standard error of the incidence requires more sophisticated concepts of error propagation, for instance, influence functions [Was06] or Taylor series approximations of random variables [Wol07]. Another way of generalizing the findings in this paper refers to the functional form of the prevalence $p$. This may be done to fit more complex prevalence data, e.g., time-age-interactions. In such cases, writing $p(t, a)$ in terms of the expit-function $p(t, a) = \text{expit}(f(t, a))$ with any differentiable function $f$, e.g., a polynomial in $t$ and $a$, immediately yields the plug-in estimate for the incidence $i = p\,(\partial f + m_1 - m_0)$ with $\partial f = \partial f/\partial t + \partial f/\partial a$. This follows from the fact that the derivative of the expit function is given by expit' $= (1 - \text{expit}) \cdot \text{expit}$ and inserting $p = \text{expit}(f)$ into Eq. (1).

An important generalization can be seen in non-chronic conditions. So far, we considered diseases without remission only. If there is a way back from the *Diseased* state to *Normal* in Figure 1 with remission rate $r$, the PDE (1) becomes [Bri15b]:

$$\left(\partial/\partial t + \partial/\partial a\right) p = (1 - p)\left\{i - p\left(m_1 - m_0\right)\right\} - r\,p\,. \tag{11}$$

In case of non-differential mortality, PDE (11) is linear and can be solved analytically, similarly to the example about the coal miners. If $m_1 \neq m_0$, an analysis similar to the diabetes example is possible. However, in both cases, differential and non-differential mortality, we need additional information (or assumptions) about the remission rate $r$ to make inference about the incidence rate $i$. Although remission from yellow fever is possible, Muench in the introductory example made the implicit assumption that a positive serostatus does not change after an infection. A similar consideration can be done in other non-chronic conditions.

# Appendices

## *Funding statement*

The authors did not receive any funding for any aspect of this work.

## *Competing interests*

The author declares that no competing interests exist with any aspect of this work.